\documentclass[twocolumn,amsmath,amssymb,aps, prl]{revtex4-2}
\usepackage{graphicx}
\usepackage{color,soul}
\usepackage{amsmath,amssymb}
\usepackage{mathrsfs}
\usepackage{color}
\usepackage{afterpage}
\usepackage[version=3]{mhchem}
\usepackage{natbib}
\usepackage{soul}
\usepackage[caption=false]{subfig}
\usepackage{array}
\usepackage{multirow}
\renewcommand{\thesection}{\Roman{section}} 
\usepackage[colorlinks, citecolor=blue,urlcolor=blue, linkcolor=blue, bookmarks=false]{hyperref}
\hypersetup{colorlinks=true , citecolor=blue, urlcolor=blue, linkcolor=blue}
\begin{document}

\title{Multiple Zeeman-type Hidden Spin Splittings in $\mathcal{\hat{P}\hat{T}}$-Symmetric Layered Antiferromagnets}

\author{Sajjan Sheoran} 
\email{phz198687@physics.iitd.ac.in}
\author{Saswata Bhattacharya}
\email{saswata@physics.iitd.ac.in}
\affiliation{Department of Physics, Indian Institute of Technology Delhi, New Delhi 110016, India}

\begin{abstract}
  Centrosymmetric antiferromagnetic semiconductors, although abundant in nature, appear less favorable in spintronics owing to the lack of inherent spin polarization and magnetization. We unveil hidden Zeeman-type spin splitting (HZSS) in layered centrosymmetric antiferromagnets with asymmetric sublayer structures by employing first-principles simulations and symmetry analysis. Taking the bilayer counterpart of recently synthesized monolayer MnSe, we demonstrate that the degenerate states around specific high-symmetry points spatially segregate on different sublayers forming $\mathcal{\hat{P}\hat{T}}$-symmetric pair. Furthermore, degenerate states exhibit uniform in-plane spin configurations with opposite orientations enforced by mirror symmetry. Bands are locally Zeeman-split up to order of $\sim$70 meV. Strikingly, a tiny electric field of a few mV{\AA$^{-1}$ along the $z$-direction breaks the double degeneracy forming additional Zeeman pair}. Moreover, our simulations on trilayer and tetralayer MnSe show that achieved HZSS is independent of layer number. These findings establish the design principle to obtain Zeeman-type splitting in centrosymmetric antiferromagnets and significantly expand the range of materials to look for spintronics.
\end{abstract}
\maketitle

Spin polarization in nonmagnetic crystals without inversion symmetry ($\mathcal{\hat{P}}$) can be achieved through relativistic spin-orbit coupling (SOC), as demonstrated by Dresselhaus~\cite{dresselhaus1955spin} and Rashba~\cite{rashba1960properties,vas1979spin, bychkov1984properties} in their influential works. Recent research has indicated that a similar phenomenon called hidden spin polarization (HSP) can exist even in centrosymmetric crystals, provided that individual atomic sites break local inversion symmetry~\cite{zhang2014hidden,bertoni2016generation, yao2017direct,yuan2019uncovering,huang2020hidden,clark2022hidden}. The discovery of HSP opens up possibilities for a broader range of materials in spintronics and offers new insights into various hidden physical properties such as optical polarization~\cite{qian2016coherence,liu2015intrinsic}, valley polarization~\cite{liu2015intrinsic}, orbital polarization~\cite{ryoo2017hidden,beaulieu2020revealing}, and Berry curvature~\cite{cho2018experimental,schuler2020local}. The HSP effect in nonmagnetic materials, characterized by the coexistence of $\mathcal{\hat{P}}$ and time-reversal symmetry ($\mathcal{\hat{T}}$), is odd distributed in both real and momentum spaces within a localized sector~\cite{liu2015search}. While techniques like spin- and angle-resolved photoemission spectroscopy have successfully measured HSP with both positional (\textbf{\textit{r}}) and momentum (\textbf{\textit{k}}) resolution, its application in spintronics necessitates the breaking of global symmetry, typically achieved through external electric fields~\cite{yuan2019uncovering,liu2015intrinsic}.

Antiferromagnets have recently emerged as a viable substitute for nonmagnetic and ferromagnetic materials in spintronic applications~\cite{vzelezny2018spin,nvemec2018antiferromagnetic, yuan2021prediction, yuan2021strong, he2023nonrelativistic}. Due to their resilience against magnetic disruptions, lack of stray fields, and ability to exhibit exceptionally rapid spin dynamics, antiferromagnets possess the potential to outperform ferromagnets. Recent research focused on investigating spin splittings within various magnetic space groups, encompassing antiferromagnets~\cite{yuan2023degeneracy,bai2022observation}. However, centrosymmetric antiferromagnetic (AFM) semiconductors, which lack polarization and have either negligible or nonexistent magnetization, pose challenges in generating substantial and controllable spin splitting using magnetic or electric fields~\cite{sivadas2016gate,li2023progress, zhao2022zeeman}. {As a result, despite their abundance in nature, centrosymmetric AFM semiconductors do not appear promising for practical applications in spintronics. Interestingly, the magnetoelectric effect~\cite{fiebig2005revival}}, in which an electric field has a dual effect of inducing both polarization and magnetization, allows control over spin splitting and textures in centrosymmetric antiferromagnets even without SOC~\cite{zhao2022zeeman}. The gate-controlled Magneto-optic Kerr effect in layered collinear AFM MnPSe$_3$~\cite{sivadas2016gate} and anomalous Hall effect in orthorhombic AFM CuMnAs~\cite{cao2023plane} are observed. Additionally, two AFM semiconductors, Fe$_2$TeO$_6$ and SrFe$_2$S$_2$O, show spin-splittings as large as 55 and 30 meV induced by the electric field of strength {60 mV/\AA~\cite{zhao2022zeeman}}. Even though the magnetoelectric coupling effect is observed in several centrosymmetric antiferromagnets through first-principles simulations~\cite{sivadas2016gate,li2023progress, zhao2022zeeman,cao2023plane}, the attainment of HSP remains uncommon due to the stringent requirements imposed on lattice and site symmetry.
\begin{figure}[htp]
    \includegraphics[width=8.6cm]{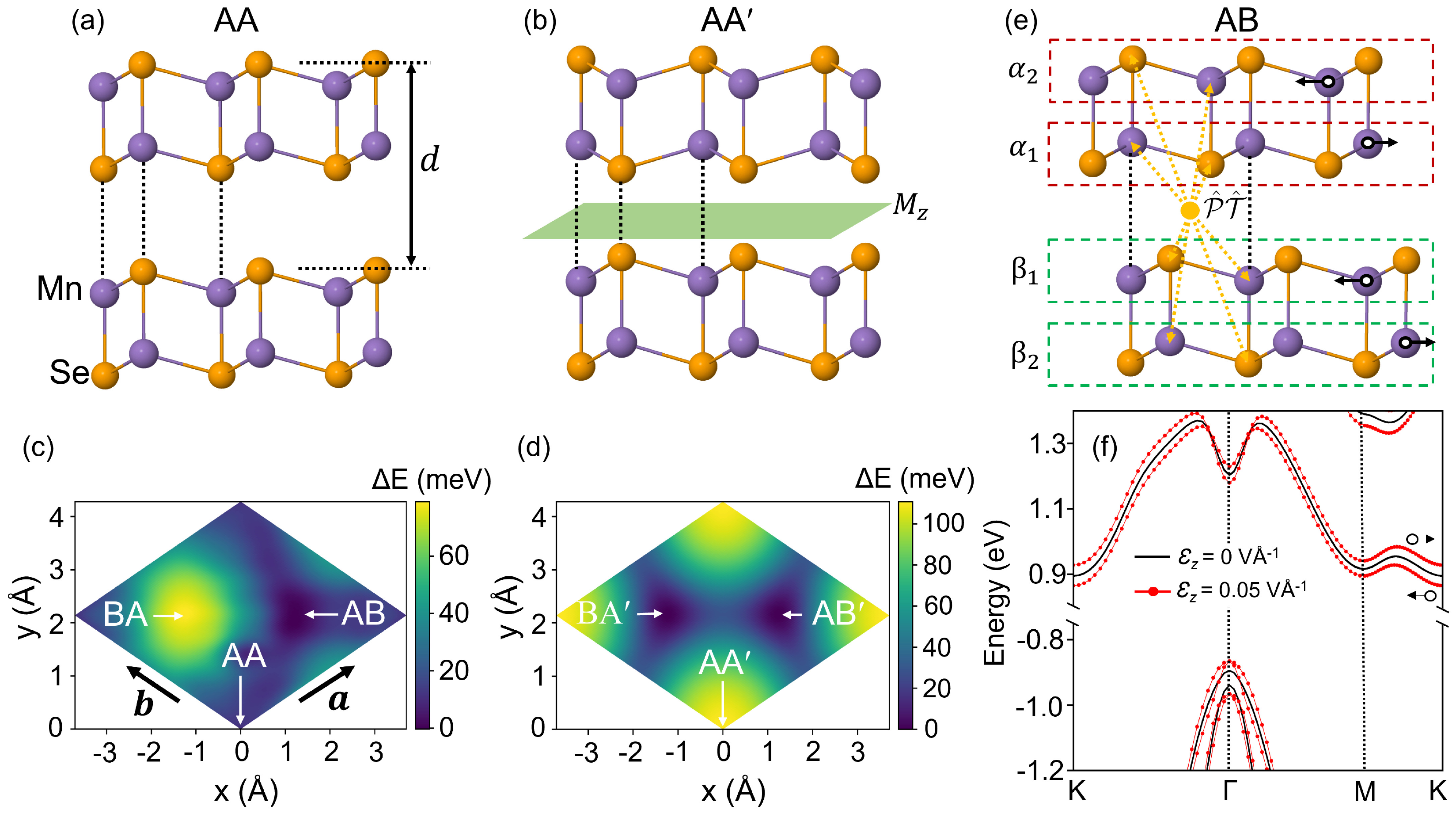}
	\caption{Geometric structures of bilayer MnSe with high-symmetry stackings (a) AA and (b) AA$'$. The upper layer in AA and AA$'$ is obtained by the translation ($d\hat{z}$) and mirror reflection ($M_z$) of the lower layer, respectively. The energy distribution of different translations between layers for the (c) AA and (d) AA$'$ stacking. The energy surfaces are interpolated using 10 steps in {\textbf{\textit{a}} and \textbf{\textit{b}}} directions such that all possible high-symmetry stackings are included. (e) Structure of the global ground state with AB stacking having magnetization $M$$\uparrow \downarrow \uparrow \downarrow$. The sublayers $\alpha_i$ and $\beta_i$ ($i=1,2$) are connected by $\mathcal{\hat{P}\hat{T}}$-center with orange dashed lines. (f) Energy band dispersion of monolayer MnSe without (solid black lines) and with an electric field (red dashed lines) of 0.05 V\AA$^{-1}$ along $z$-direction. Arrows are used to illustrate spin orientations.}
	\label{p1}
\end{figure}

In this letter, we examine the occurrence of hidden Zeeman-type spin splitting (HZSS) in AFM 2D van der Waals (vDW) bilayers and multilayers made up of the newly synthesized monolayer MnSe~\cite{aapro2021synthesis} using a combination of symmetry analysis and density-functional theory (DFT) calculations. Scanning potential energy surfaces of all possible high-symmetry stacking, we show that bilayer MnSe with $\mathcal{\hat{P}\hat{T}}$ symmetry is energetically most favorable. $\mathcal{\hat{P}\hat{T}}$ symmetry leads to the HZSS with minimal mixing between degenerate states segregated on different sublayers. In addition, the mirror plane ($\hat{M}_x$) enforces the spins to be \textbf{\textit{k}}-independent. Strikingly, a tiny electric field splits the otherwise doubly degenerate bands preserving the spin configurations. {Switching the electric field can reverse the electronic spin magnetization connected to the energy level splitting.} Afterward, we extend our approach to the trilayer and tetralayer MnSe to show the layer number independence of our observations.

The MnSe monolayer consists of two buckled honeycomb sublayers of MnSe, interconnected by Mn-Se bonds {[Fig.~\ref{p1}a]}. Within the monolayer, the upper Mn/Se atoms are positioned alternately on top of the lower Se/Mn atoms. Each unit cell of monolayer contains {two Mn and two Se atoms}, with a lattice constant of 4.28 Å, belonging to the space group $P\overline{3}m1$ with $\mathcal{P}$ symmetry (ignoring spin configurations). Considering different possible magnetic configurations, we find that the upper and lower sublayers are coupled antiferromagnetically through Neel-type antiferromagnetism (see section I of the supplemental Material (SM)~\cite{SuMa} for computational methods). The robust Neel-type AFM state is more stable than the FM state by 0.38 eV per unit cell (u.c.). Additionally, the magnetic anisotropic energy (MAE) calculations reveal that the monolayer MnSe prefers in-plane magnetization over out-of-plane magnetization, with an energy difference of 0.37 meV/u.c. These results are consistent with the previous experimental~\cite{aapro2021synthesis} and theoretical reports~\cite{aapro2021synthesis,sattar2022monolayer,liu2023tunable}.

Determination of stacking orders in multilayers is challenging, especially for more complex structures. Firstly, we have created bilayer MnSe by taking the upper layer as translation (by $d$$\hat{z}$) and mirror reflection ($\hat{M}_z$) of the lower layer, as shown in Figs.~\ref{p1}(a) and \ref{p1}(b). We name these stacking AA (obtained through $d$$\hat{z}$) and AA$'$ (obtained through $\hat{M}_z$), following the nomenclature used in Ref.~\cite{2013stacking}. After that, we transformed these stackings into other possible high-symmetry stacking by translation sliding of one of the basal planes in the unit cell: going to AB and BA state from AA, and going to AB$'$ and BA$'$ from AA$'$. Further, DFT calculations were conducted to ascertain the lowest energy states for various stacking arrangements. As shown in Fig.~\ref{p1}(c), the top layer sliding from AA to AB results in a nondegenerate global energy minimum, whereas BA stacking is energetically unfavorable. On the other hand, shifting the top layer from AA$'$ leads to doubly degenerate energy minima AB$'$ and BA$'$ [Fig.~\ref{p1}(d)]. AB stacking is energetically most favorable, having lower energy than AB$'$/BA$'$ by 14 meV per unit cell. Bilayer MnSe can be divided into sublayers $\alpha_i$ and $\beta_i$ ($i=$1,2), each sublayer containing a single Mn atom. Note that we have taken four nondegenerate magnetic configurations $M$$\uparrow \uparrow \uparrow \uparrow$, $M$$\uparrow \downarrow \uparrow \downarrow$, $M$$\uparrow \downarrow \downarrow \uparrow$, and $M$$\uparrow \uparrow \downarrow \downarrow$ (here, up and down arrows represent the magnetic moment direction of sublayers along $+x$ and $-x$ direction, respectively, in the order of $\beta_2$, $\beta_1$, $\alpha_1$, and $\alpha_2$ [Fig.~\ref{p1}(e)]). Given that the easy axis of magnetization lies in the $x$-$y$ plane with MAE of 0.9 meV/u.c., we have chosen the $x$-direction as the axis for magnetization. Our DFT calculations discover that Neel-type AFM configuration $M$$\uparrow \downarrow \uparrow \downarrow$ has the lowest energy. Therefore, bilayer MnSe is most likely to exist in AB stacking with $M$$\uparrow \downarrow \uparrow \downarrow$ configuration, which is obtained by translating the top layer in AA by $\frac{2}{3}\textbf{\textit{a}}+\frac{1}{3}\textbf{\textit{b}}$. Our calculation assumes that the bilayer MnSe exhibits AB stacking with $M$$\uparrow \downarrow \uparrow \downarrow$ configuration unless otherwise specified.

 \begin{table}[h]
  	\caption{The transformation rules of ($P_x, P_y, P_z$) and ($\hat{\sigma}_{x}, \hat{\sigma}_{y}, \hat{\sigma}_{z}$) with respect to the operations ($\hat{O}$) belonging to MPG $2'/m$. The last column shows the terms which are invariant under operation.}
		\centering
		\label{t1}
		\begin{tabular}{c c c c }
			\hline
			\hline
			 $\hat{O}$ & ($P_x, P_y, P_z$) & ($\hat{\sigma}_{x}, \hat{\sigma}_{y}, \hat{\sigma}_{z}$) & Invariants \\ \hline
			 $\hat{I}$ & ($P_x, P_y, P_z$) & ($\hat{\sigma}_{x}, \hat{\sigma}_{y}, \hat{\sigma}_{z}$) & $P_i\hat{\sigma}_j(i,j$=$x,y,z)$ \\
			 $\hat{M}_x$ & ($-P_x, P_y, P_z$) & ($\hat{\sigma}_{x},-\hat{\sigma}_{y},-\hat{\sigma}_{z}$) & $P_x \hat{\sigma}_{y/z}$, $P_{y/z}\hat{\sigma}_{x}$ \\
			 $\hat{C}_{2x}$$\mathcal{\hat{T}}$ & ($P_x, -P_y, -P_z$) & ($-\hat{\sigma}_{x}, \hat{\sigma}_{y}, \hat{\sigma}_{z}$) & $P_x \hat{\sigma}_{y/z}$, $P_{y/z}\hat{\sigma}_{x}$\\
			 $\mathcal{\hat{P}\hat{T}}$ & ($-P_x, -P_y, -P_z$) & ($-\hat{\sigma}_{x}, -\hat{\sigma}_{y}, -\hat{\sigma}_{z}$) & $P_i\hat{\sigma}_j(i,j$=$x,y,z)$ \\
			\hline
			 \hline
		\end{tabular}
\end{table}

\begin{figure}[htp]
	\includegraphics[width=8.5cm]{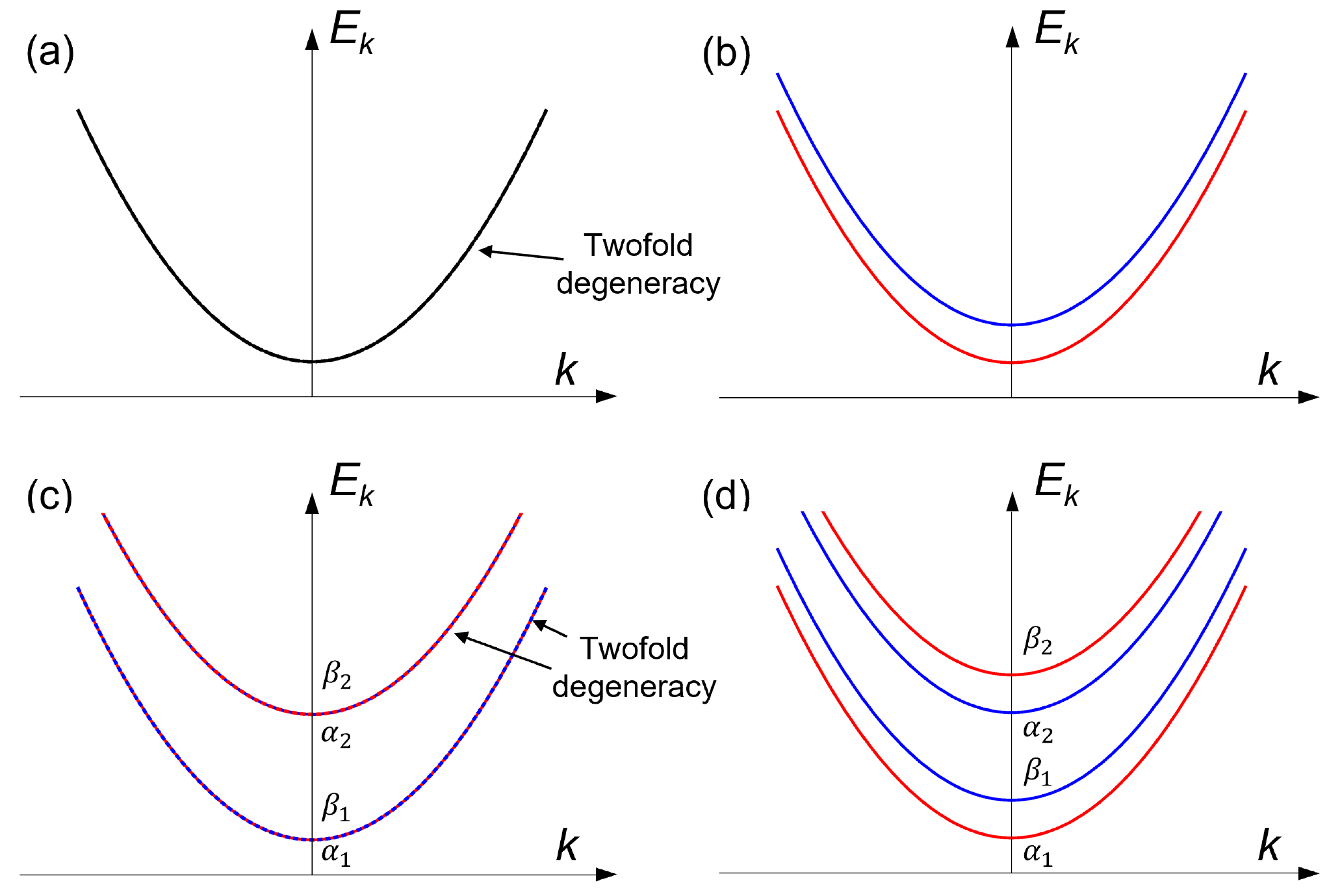}
	\caption{Schematic illustration of band structures with (a) null spin splitting and (b) Zeeman spin splitting. (c) Sketches of twofold degenerate bands in the presence of $\hat{P}\hat{T}$ symmetry with hidden spin polarization, where each band is segregated on different sublayers. (d) The lifting of band degeneracy in the presence of an electric field along the $z$-direction. The energy levels with positive and negative spin magnetization are sketched in the red and blue curves, respectively.}
	\label{p2}
\end{figure}

The magnetic point group (MPG) for bilayer MnSe with AB stacking is $2'/m$ having symmetry operations $\hat{M}_x$, $\hat{C}_{2x}$$\mathcal{\hat{T}}$, and $\mathcal{\hat{P}\hat{T}}$ besides trivial identity operation ($\hat{I}$). Where $\hat{M}_x$ and $\hat{C}_{2x}$ are mirror reflection in the $y$-$z$ plane and two-fold rotation around the $x$-axis, respectively.  $\alpha_i$ is connected to the $\beta_i$ through  $\mathcal{\hat{P}\hat{T}}$ symmetry {[Fig.~\ref{p1}(e)]}. Therefore, we say that $\alpha_i$ is the inversion partner of $\beta_i$ and vice versa. Due to the $\mathcal{\hat{P}\hat{T}}$ symmetry, spin polarization $S_{\textbf{\textit{k}}}(\textbf{\textit{r}})$ is odd distributed in both \textbf{\textit{r}} and \textbf{\textit{k}} spaces, i.e. $S_{\textbf{\textit{k}}}^{\alpha_i}$=$-S_{\textbf{\textit{k}}}^{\beta_i}$ and $S_{\textbf{\textit{k}}}^{\alpha_i(\beta_i)}$=$-S_{-\textbf{\textit{k}}}^{\alpha_i(\beta_i)}$~\cite{chen2022role}. Furthermore, the  energy of state localized on $\alpha_i$ sublayer $|\psi_{\textbf{\textit{k}},\uparrow}^{\alpha_i} \rangle $ ($E_{\textbf{\textit{k}}, \uparrow}^{\alpha_i}$) and state localized on $\beta_i$ sublayer $| \psi_{\textbf{\textit{k}},\downarrow}^{\beta_i} \rangle $ ($E_{\textbf{\textit{k}}, \downarrow}^{\beta_i}$) are degenerate [Fig.~\ref{p2}]. Although energy bands are degenerate, the two components of doubly degenerate bands have opposite spin polarization, each spatially localized on one of the two separate sublayers forming the inversion partners called the spin-sublayer locking effect [Fig.~\ref{p2}(c)].  While the total spin polarization is prohibited by $\mathcal{\hat{P}\hat{T}}$ symmetry, it is possible to observe local spin polarization with opposite values in sublayers $\alpha_i$ and $\beta_i$. When examining a specific layer (say $\alpha$) that consists of sublayers $\alpha_1$ and $\alpha_2$, the states localized on $\alpha_1$ and $\alpha_2$ have non-degenerate energy. Consequently, the states $|\psi_{\textbf{\textit{k}},\uparrow}^{\alpha_1} \rangle $ and $|\psi_{\textbf{\textit{k}},\downarrow}^{\alpha_2} \rangle $ create a Zeeman pair. Similarly, $\beta_1$ and $\beta_2$ states also make a Zeeman Pair. Moreover, the anticommutation relationship between $\hat{M}_x$ and $\hat{\sigma}_{y,z}$ in spin space (where $\hat{\sigma}_{x,y,z}$ represents the Pauli matrices) imposes a condition on the general energy eigenstate $|\psi_{\textbf{\textit{k}},\uparrow/\downarrow}^{\alpha_i/\beta_i} \rangle $. Specifically, $\langle \psi_{\textbf{\textit{k}},\uparrow/\downarrow}^{\alpha_i/\beta_i} | \hat{M}_x^{-1} \hat{\sigma}_{y,z} \hat{M}_x | \psi_{\textbf{\textit{k}},\uparrow/\downarrow}^{\alpha_i/\beta_i} \rangle $=$-\langle \psi_{\textbf{\textit{k}},\uparrow/\downarrow}^{\alpha_i/\beta_i} |\hat{\sigma}_{y,z}| \psi_{\textbf{\textit{k}},\uparrow, \downarrow}^{\alpha_i/\beta_i} \rangle $, which implies that $\langle \psi_{\textbf{\textit{k}},\uparrow/\downarrow}^{\alpha_i/\beta_i} |\hat{\sigma}_{y,z}| \psi_{\textbf{\textit{k}},\uparrow/\downarrow}^{\alpha_i/\beta_i} \rangle $=0. Consequently, the spin orientations are independent of \textbf{\textit{k}} and aligned parallel or antiparallel to $x$ direction. That leads to the HZSS with persistent spin textures [Fig.~\ref{p2}(c)], and energy relationship can be expressed as
\begin{equation}
	E_{\textbf{\textit{k}}, \uparrow}^{\alpha_i}=E_{\textbf{\textit{k}}, \downarrow}^{\beta_i} ; i=1,2
	\quad\text{and}\quad 
	E_{\textbf{\textit{k}}, \uparrow}^{s_1}\neq E_{\textbf{\textit{k}}, \downarrow}^{s_2}  ; s=\alpha,\beta
	\label{e1}
\end{equation}

Next, we apply an {electric field (\textbf{\textit{$\mathcal{E}$}}) to break $\mathcal{\hat{P}\hat{T}}$ symmetry. The electric field creates a net polarization ($\textbf{\textit{P}}\propto$ \textbf{\textit{$\mathcal{E}$}})~\cite{fiebig2005revival}}. Induced polarization generates a net magnetization ($M_i\propto P_j$) through the magnetoelectric effect~\cite{fiebig2005revival}, lifting the $\mathcal{\hat{P}\hat{T}}$ {symmetry} between inversion partners. The occurrence of magnetization induces an effective magnetic field ($\textbf{\textit{B}}^{eff}\propto \textbf{\textit{P}}$) that couples with the spin degrees of freedom, yielding a Zeeman-like Hamiltonian $\sum_{i,j}\lambda_{i,j}P_i\hat{\sigma}_j$. Tensor $\lambda_{i,j}$ determines the strength of the magnetoelectric coupling. Some components of $\lambda_{i,j}$ are never non-zero and vanish due to the symmetry. We determine the symmetry-allowed terms $\lambda_{ij}$ using the method of invariants~\cite{voon2009kp} ($\hat{H} = \hat{O}^{\dagger}\hat{H}\hat{O}$, where $\hat{O}$ is the symmetry operation belonging to the MPG $2'/m$), generally used to determine the \textbf{\textit{k.p}} Hamiltonian in nonmagnetic materials~\cite{tao2017reversible,sheoran2023coupled}. The Zeeman Hamiltonian ($\hat{H}_Z$), following the transformation rules of $P_i$ and $\hat{\sigma}_j$ listed in Table~\ref{t1}, is given by
\begin{equation}
	\hat{H}_Z= \lambda_{x,y} P_x \hat{\sigma}_{y} + \lambda_{x,z} P_x \hat{\sigma}_{z}+ \lambda_{y,x} P_y \hat{\sigma}_{x}+\lambda_{z,x} P_z \hat{\sigma}_{x}.
	\label{e2}
\end{equation}
$\hat{H}_Z$  breaks the $\mathcal{\hat{P}\hat{T}}$ symmetry and lifts the energy degeneracy between states localized on inversion partners [Fig.~\ref{p2}(d)] and thus forming additional Zeeman-like pairs
\begin{equation}
	E_{\textbf{\textit{k}}, \uparrow}^{\alpha_i}\neq E_{\textbf{\textit{k}}, \downarrow}^{\beta_i} ; i=1,2
	\quad\text{and}\quad 
	E_{\textbf{\textit{k}}, \uparrow}^{s_1}\neq E_{\textbf{\textit{k}}, \downarrow}^{s_2}  ; s=\alpha,\beta.
	\label{e3}
\end{equation}
 It should be noted that $\hat{H}_Z$ is determined assuming that the magnetization axis aligns with the $x$-direction. However, the Hamiltonian can be adapted for any in-plane direction by rotating it accordingly.

\begin{figure}[htp]
	\includegraphics[width=8.6cm]{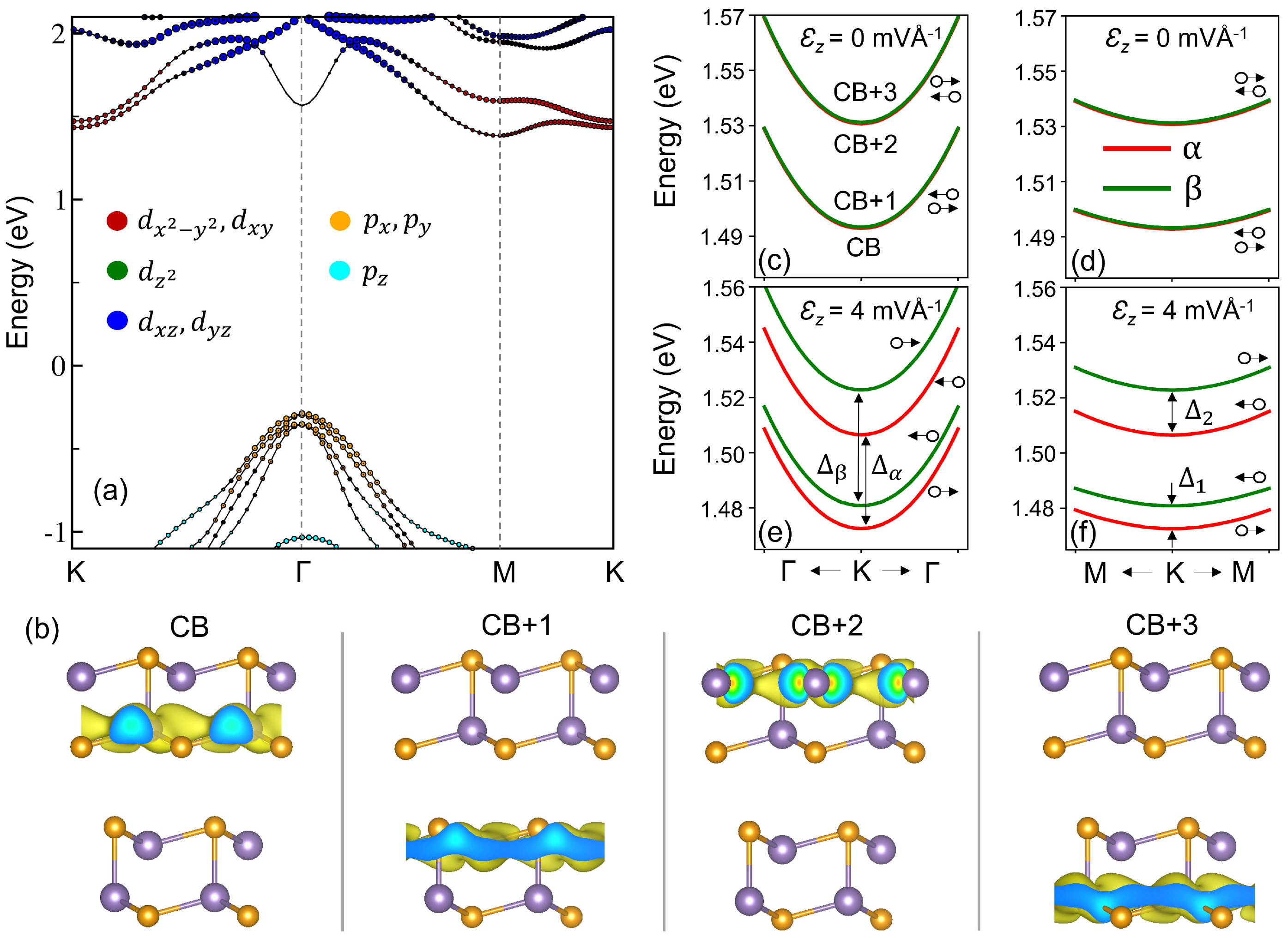}
	\caption{(a) Orbital-projected band dispersion curves of bilayer MnSe with AB stacking. (b) Charge density distribution plotted for the states CB, CB+1, CB+2, and CB+3 at K point. Layer-projected CBs of bilayer MnSe around the K point along the (c) K-$\Gamma$ and (d) K-M. (e)-(f) counterparts of (c)-(d) in the presence of a small electric field of 4 mV{\AA}$^{-1}$ along the $z$-direction.}
	\label{p3}
\end{figure}

 According to symmetry analysis, there are two conditions to observe HZSS: (i) $\hat{P}\hat{T}$ symmetry and (ii) spin-sublayer locking effect. Condition (i) is already fulfilled, and to check condition (ii), we plot the orbital-projected band structure of bilayer MnSe considering spin-orbit interaction [Fig.~\ref{p3}(a)]. Under $C_{3v}$ site symmetry, the Mn-$d$ orbitals split into three categories $d_{z^2}$, $(d_{x^2-y^2},d_{xy})$, and $(d_{xz},d_{yz})$. The Se-$p$ orbitals are split into two categories $p_z$ and $(p_x$, $p_y)$. The Mn-$d$ orbitals form CB states, while Se-$p$ orbitals mainly contribute to VB states. Compared to the band structure of monolayer MnSe [Fig.~\ref{p1}(f)], interlayer interaction between inversion partners $\alpha_1$ and $\alpha_2$ results in increased separation between the nested bands and spin mixture from another sublayer. Compared to the in-plane orbitals ($d_{x^2-y^2}, d_{xy}, p_x, p_y$), the out-of-plane orbitals ($d_{xz}, d_{yz}, d_{z^2}, p_z$) exhibit significant interlayer interaction. Observing HZSS is not ideal with the CB states around the M and $\Gamma$ points and {the VB states around the K and M points}. That is because these states originate from the hybridization of out-of-plane orbitals between different sublayers, and thus their wavefunctions are not segregated. The CB states near the K point are predominantly contributed by $(d_{x^2-y^2},d_{xy})$ and VB states around the $\Gamma$ point are composed of mainly $(p_x$, $p_y)$. Therefore states arising from the in-plane orbitals, {i.e., the CBs near the K point and the VBs near the $\Gamma$ point}, have weak communication between different sublayers, leading to minimal separation between nested bands and spin mixture from neighboring layers. Consequently, each wavefunction is localized within the sublayer for the CBs near the K point and the {VBs near the $\Gamma$ point}, forming the ideal choice to observe spin-sublayer locking effect and HZSS [Fig.~\ref{p3}(b)]. 

Next, we plot layer-projected dispersion curves for low CBs in the vicinity of K point for bilayer MnSe [Figs.~\ref{p3}(c) and~\ref{p3}(d)]. The four lowest CBs form two two-fold degenerate pairs [see Eq.~\ref{e1}]. Despite having the same energy, these states are localized on distinct sublayers and can be probed using \textbf{\textit{r}}-resolved spectroscopy techniques~\cite{yao2017direct,lu2023unlocking}. For instance, the degenerate states CB and CB+1 segregate on inversion partners $\alpha_1$ and $\beta_1$, respectively [Fig.~\ref{p3}(b)]. Both inversion partners possess finite and opposite \textbf{\textit{k}}-independent spin orientations, either parallel or antiparallel to the $x$ direction [see section II in the supplemental material (SM)~\cite{SuMa}], ensuring zero net spin polarization respecting the $\mathcal{\hat{P}\hat{T}}$ symmetry requirement. The $k$-independent persistent spin textures overcome spin dephasing and provide non-dissipative spin transport~\cite{tao2018persistent,zhao2020purely,sheoran2022emergence}. $\alpha_1$ sublayer has a magnetization of magnitude 4.38 $\mu_B$ along +$x$ direction and compensated by the opposite magnetization of inversion partner $\beta_1$. A similar analysis also applies to the upper degenerate pair CB+2 and CB+3, where wavefunctions corresponding to the CB+2 and CB+3 have segregated {on} $\alpha_2$ and $\beta_2$ sublayers, respectively. If we focus locally on each layer, ($|\psi_{K,\uparrow}^{\alpha_1} \rangle $, $|\psi_{K,\downarrow}^{\alpha_2} \rangle $) and ($|\psi_{K,\downarrow}^{\beta_1} \rangle $, $|\psi_{K,\uparrow}^{\beta_2} \rangle $) form two Zeeman pair with energy splitting $\Delta_{\alpha}$, $\Delta_{\beta}$, respectively {[Figs.~\ref{p3}(c) and~\ref{p3}(d)]}. As a consequence of $\mathcal{\hat{P}\hat{T}}$ symmetry, the strength of the Zeeman spin splitting is the same for each layer ($\Delta_{\alpha}$=$\Delta_{\beta}$) to satisfy the Kramers' degeneracy. We have observed Zeeman spin splitting of order $\sim$39.5 meV. {Observed Zeeman splitting in our case (without external fields)} is comparable to bulk AFM semiconductors Fe$_2$TeO$_6$ (55 meV) and SrFe$_2$S$_2$O (33 meV) under the external electric field of strength 60 mV\r{A}$^{-1}$~\cite{zhao2022zeeman}. However, it is still smaller than nonmagnetic transition metal dichalcogenides, i.e., MoS$_2$ (150 meV)~\cite{xiao2012coupled}. {HZSS is also observed for the VB states near the $\Gamma$ point,} where both $\alpha$ and $\beta$ bands are {Z}eeman split ($\Delta_{\alpha}$, $\Delta_{\beta}$) by $\sim$70 meV, as discussed in section III of SM~\cite{SuMa}.

\begin{figure}[htp]
	\includegraphics[width=8.6cm]{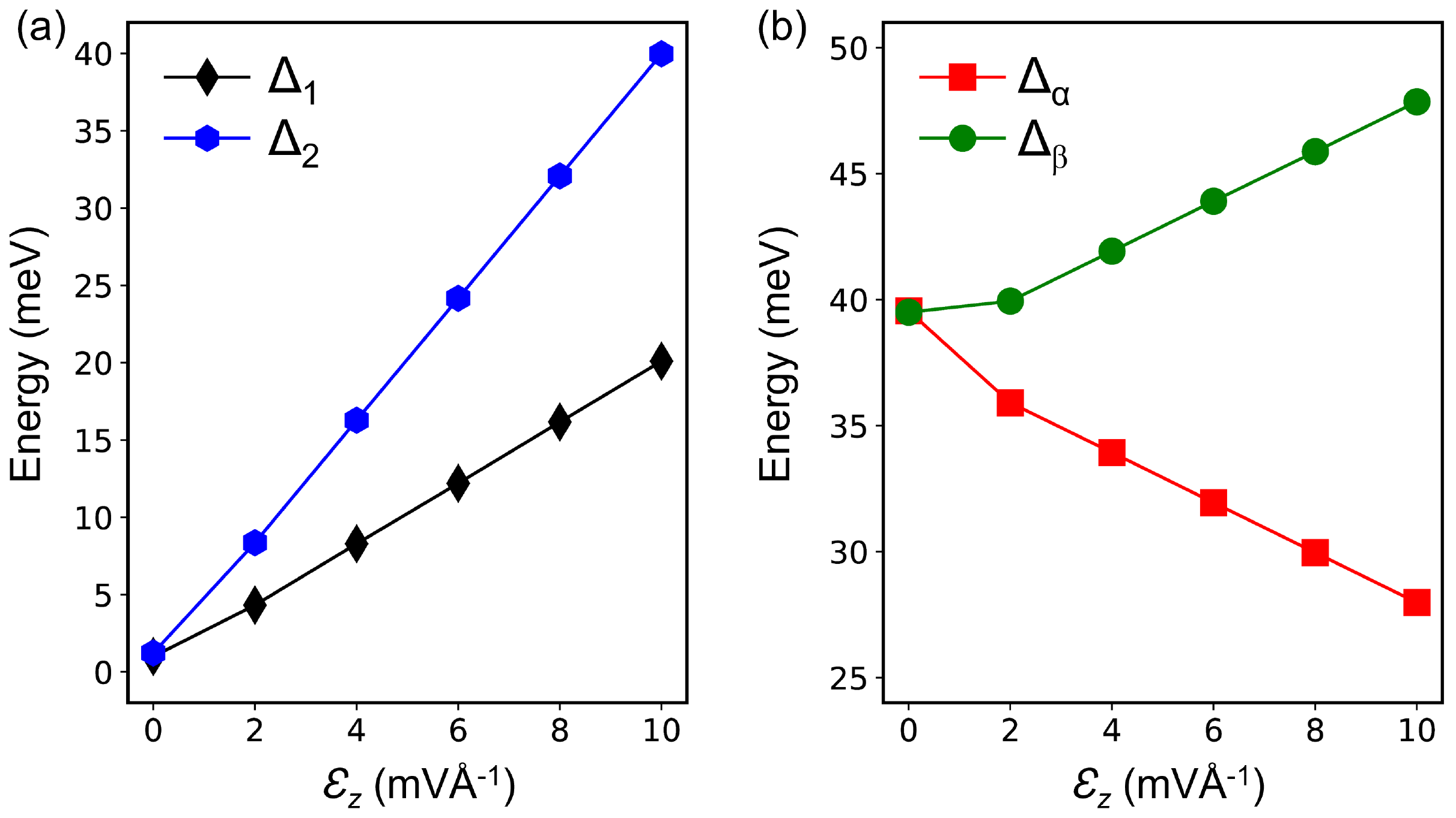}
	\caption{The Zeeman spin splittings for the CBs (a) $\Delta_{1}$, $\Delta_{2}$ and (b) $\Delta_{\alpha}$, $\Delta_{\beta}$ in bilayer MnSe as a function of electric field $\mathcal{E}_z$.}
	\label{p4}
\end{figure}

The $\hat{H}_Z$ expressed in Eq.~\ref{e2} shows that two-fold degeneracy will be lifted by including polarization along an arbitrary direction ($P_x, P_y, P_z$). We include $P_z$ through the application of an electric field along the $z$ direction ($\mathcal{E}_z$), and our model in Eq.~\ref{e2} can readily be generalized to
\begin{equation}
	\hat{H}_Z=\gamma_{z,x} \mathcal{E}_z\hat{\sigma}_{x}
	\label{e4}
\end{equation}
where $\gamma_{z,x}$ determines the strength of splitting. We plot the {CBs} in a tiny {$\mathcal{E}_z$ of 4 mV{\AA$^{-1}$} [Figs.~\ref{p3}(e) and~\ref{p3}(f)]}. The $\mathcal{E}_z$ creates an asymmetric potential on the inversion partners sublayers, resulting in global $\mathcal{\hat{P}\hat{T}}$ symmetry breaking. {Each degenerate pair splits into two levels $E_+=\gamma_{z,x}\mathcal{E}_z$ (eigenstate being $|+\rangle$) and $E_-=-\gamma_{z,x}\mathcal{E}_z$ (eigenstate being $|-\rangle$), where $\hat{\sigma}_{x}|+\rangle=+|+\rangle$ and $\hat{\sigma}_{x}|-\rangle=-|-\rangle$. Therefore, two degenerate pairs in bilayer MnSe splits into four energy levels, as represented by the $\Delta_1$ and $\Delta_2$ in Fig.~\ref{p3}(f). Here $\Delta_1$ ($\Delta_2$) represents the energy difference between the state localized on the $\alpha_1$ ($\alpha_2$), and the state localized on $\beta_1$ ($\beta_2$).} Upon initial observation, it is evident that $\Delta_1$$<$$\Delta_2$ due to the smaller real space separation between $\alpha_1$ and $\beta_1$ compared to that between $\alpha_2$ and $\beta_2$. Consequently, this results in a lesser potential difference between $\alpha_1$ and $\beta_1$, leading to a smaller energy splitting $\Delta_1$ compared to $\Delta_2$.

After that, we plot energy splittings ($\Delta_{1}, \Delta_{2}, \Delta_{\alpha}$, and $\Delta_{\beta}$) as a function of the $\mathcal{E}_z$ [Figs.~\ref{p4}(a) and~\ref{p4}(b)]. We notice that the energy splittings show a linear dependence on the $\mathcal{E}_z$ in the range of 0 to 10 mV{\AA$^{-1}$}. Maximum $\Delta_1$ and $\Delta_{2}$ are observed to be 20.0 and 39.9 meV, respectively, with the $\mathcal{E}_z$ of strength 10 mV{\AA$^{-1}$}. The $\mathcal{E}_z$ breaks the equality between $\Delta_{\alpha}$ and $\Delta_{\beta}$, where $\Delta_{\alpha}$ decreases and  $\Delta_{\beta}$ increases with increasing strength of the $\mathcal{E}_z$. The observed strength of $\Delta_{\alpha}$ and $\Delta_{\beta}$ are 28.0 and 47.8 meV, respectively, under the influence of 10 mV{\AA$^{-1}$} $\mathcal{E}_z$. Additionally, the reversal of the electric field from $\mathcal{E}_z$ to $-\mathcal{E}_z$ will reverse the spin magnetization $S_x$ to $-S_x$ (see section IV  of SM~\cite{SuMa}). Under an electric field of the same magnitude along the $+z$ or $-z$ direction, the strength of splittings $|\Delta_1|$ and $|\Delta_2|$ is independent of the electric field direction. However, $\Delta_1$ and $\Delta_2$ will flip their sign under electric field reversal. These results are in qualitative agreement with our model Hamiltonian $\hat{H}_z$ in Eq.~\ref{e4}. If we compare the effect of the electric field on bilayer MnSe with monolayer MnSe [Fig.~\ref{p1}(f)], we find that a large electric field is required in monolayer MnSe to obtain the Zeeman splitting of the same order. More specifically, Zeeman split pair of 50.9 meV is observed with $\mathcal{E}_z$ of strength 50 mV{\AA$^{-1}$}. Also, Zeeman splitting without any external field in bilayer MnSe is an additional advantage over monolayer MnSe.

It is interesting to note that $\hat{\mathcal{P}}$$\hat{\mathcal{T}}$ symmetry remains intact with increasing layer thickness. Therefore, effects similar to bilayer MnSe are also achieved in trilayer and tetralayer MnSe, where every band is doubly degenerate, and degenerate pair of bands are localized on different sublayers connected through $\hat{\mathcal{P}}$$\hat{\mathcal{T}}$ symmetry (see section V  of SM~\cite{SuMa}). Thus, experimental realization of HZSS can be easily achieved in layered MnSe with arbitrary thickness.

In conclusion, we have shown the existence of HZSS in bilayer MnSe, combining first-principles calculations and symmetry analysis. The stacking characteristic study shows that AB with $\hat{\mathcal{P}}\hat{\mathcal{T}}$ symmetry is the lowest energy stacking. The doubly degenerate states arising due to $\hat{\mathcal{P}}\hat{\mathcal{T}}$ symmetry are segregated into different sublayers forming inversion partners. Then, we identified the local Zeeman splitting as large as $\sim$70 meV in bilayer MnSe, where states forming Zeeman pair with opposite spin orientation segregate on different sublayers within a single layer. Interestingly, a tiny out-of-plane electric field yields additional Zeeman pairs through breaking two-fold degeneracy. Mirror symmetry ($\hat{M}_x$) enforced persistence spin textures remain preserved under an out-of-plane electric field and are known for nondissipative spin transport. Moreover, the fact that our results are independent of the layer number adds credibility to the possibility of experimental observation. Controllable Zeeman spin splittings achieved using electric fields are detectable through established approaches such as optical and transport measurements commonly used in spintronics~\cite{zhang2019semiconductor,shao2023neel}. Zeeman splittings with hidden spin polarization in centrosymmetric antiferromagnets form another prospective aspect in developing semiconductor spintronics devices~\cite{chen2022role}. We anticipate that our findings will not only contribute to a deeper understanding of magnetoelectric interactions but also inspire further innovative research in the emerging fields of AFM and semiconductor spintronics~\cite{baltz2018antiferromagnetic,gomonay2018antiferromagnetic}.

S.S. acknowledges CSIR, India, for the senior research fellowship [grant no. 09/086(1432)/2019-EMR-I]. S. B. acknowledges financial support from SERB under a core research grant (grant no. CRG/2019/000647) to set up his High-Performance Computing (HPC) facility ``Veena" at IIT Delhi for computational resources.

\bibliography{ref}

\end{document}


\title{Supplemental Material for\\
	``Multiple Zeeman-type Hidden Spin Splittings in $\mathcal{\hat{P}\hat{T}}$-Symmetric Layered Antiferromagnets"}
\author{Sajjan Sheoran\footnote{phz198687@physics.iitd.ac.in} and Saswata Bhattacharya\footnote{saswata@physics.iitd.ac.in}} 
\affiliation{Department of Physics, Indian Institute of Technology Delhi, New Delhi 110016, India}
\pacs{}
\maketitle
\begin{enumerate}[\bf I.]
\item C\MakeLowercase{omputational methods}	
\item S\MakeLowercase{pin textures}
\item HZSS \MakeLowercase{for the valence bands around $\Gamma$ point}
\item E\MakeLowercase{lectric field reversal}
\item E\MakeLowercase{xtension to Multilayer}

\end{enumerate}

\section{C\MakeLowercase{omputational methods}}
Electronic structures are calculated using the DFT implemented in the Vienna \textit{ab-initio} simulation package (VASP)~\cite{kresse1996efficient}. The ion-electron interaction is described using the projector-augmented wave method~\cite{kresse1999ultrasoft}, and the exchange-correlation interaction is treated using the Perdew-Burke-Ernzerhof~\cite{perdew1996generalized} generalized gradient approximation. To discretize the Brillouin zone, we employ k mesh with spacing 0.02 {\AA$^{-1}$}, and a plane-wave expansion with a cutoff of 520 eV is utilized. Our calculations include the SOC, however the qualitative picture of Zeeman splittings remain the same even without SOC. The DFT-D3 approach was used to deal with vDW interactions~\cite{grimme2006semiempirical}. The periodic slabs are separated by a vacuum layer with a thickness greater than 15 Å. All structures undergo optimization until the residual force on each atom is below 0.001 eV/Å. We employ the GGA+U method~\cite{liechtenstein1995density} to incorporate the onsite Coulomb interaction of localized Mn-3d orbitals and set the onsite Coulomb (U) and exchange parameter (J) to 2.3 and 0 eV, respectively, as used in the previous report for monolayer MnSe~\cite{aapro2021synthesis}.  We follow the approach introduced by Neugebauer and Scheffler~\cite{neugebauer1992adsorbate} to introduce a uniform electric field to the layered MnSe slab. This method addresses the artificial periodicity of the slab by incorporating a flat dipole layer within the vacuum region. The dipole's strength is determined through self-consistent calculations, ensuring compensation for the induced dipole caused by the electrostatic field. We have conducted symmetry analysis using Ref.~\cite{dresselhaus2007group},  Bilbao crystallographic server~\cite{aroyo2006bilbao,elcoro2017double}, FINDSYM~\cite{stokes2005findsym}, and SEEK-PATH~\cite{hinuma2017band}. In addition, various packages and toolkits, including  VASPKIT~\cite{wang2021vaspkit}, PyProcar~\cite{herath2020pyprocar}, and MAGNDATA~\cite{gallego2016magndata}, were also helpful.
\section{S\MakeLowercase{pin textures}}
\begin{figure}[h]
	\includegraphics[width= 5 in]{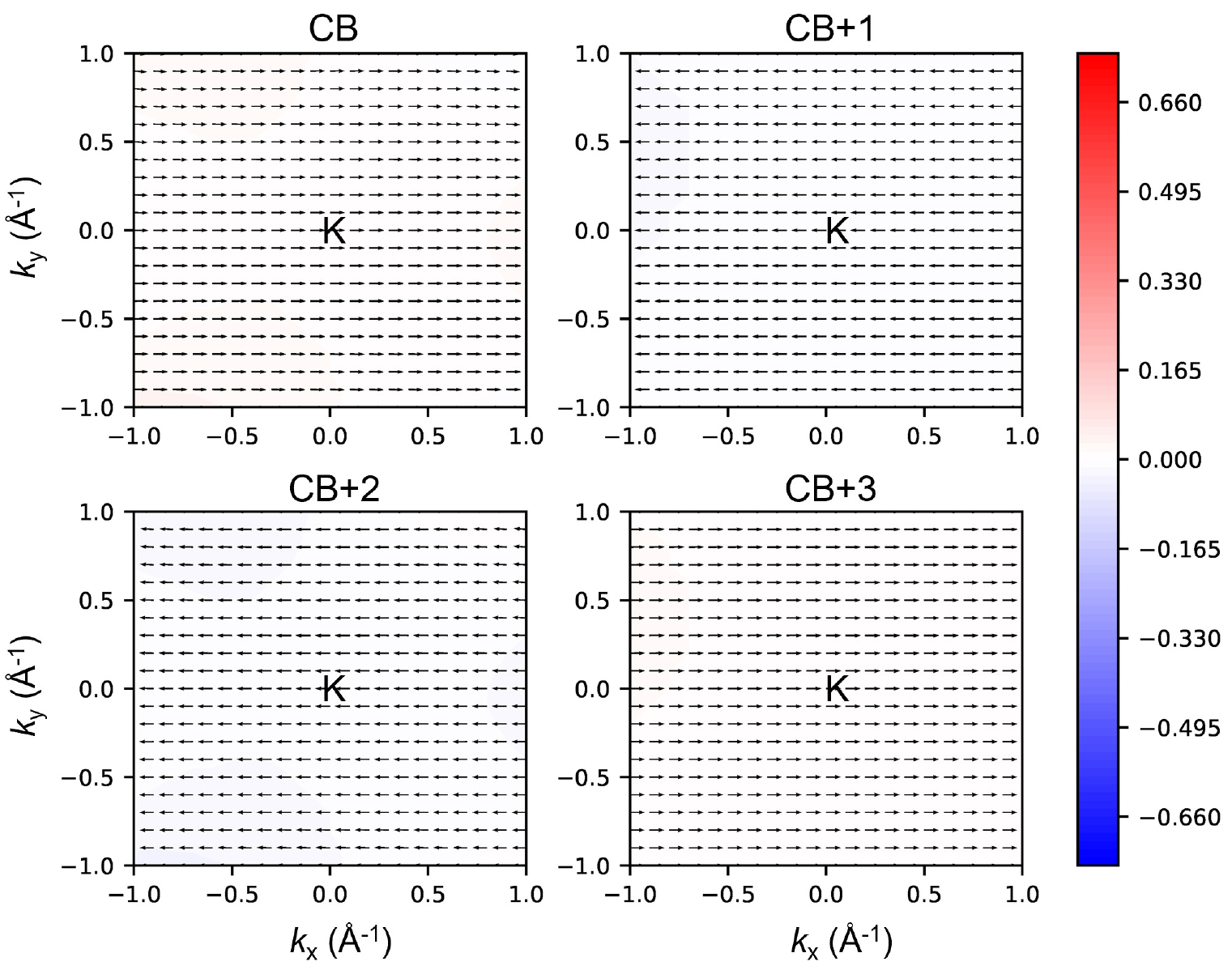}
	\caption{Spin textures of four lowermost conduction bands CB, CB+1, CB+2, and CB+3 around the K point. The arrows and color represent the in-plane ($x,y$) and out-of-plane ($z$) components of spin textures, respectively.}
	\label{s1}
\end{figure}

We use DFT to study the spin orientations as a function of momentum ($\textbf{\textit{k}}$), generally referred to as spin textures~\cite{gomonay2018antiferromagnetic}. Our primary focus is centered on the conduction bands around the K point in bilayer MnSe. We plot the spin textures using $\langle \psi_{\textbf{\textit{k}}} |\hat{\sigma}_{i}| \psi_{\textbf{\textit{k}}}\rangle$ ($i=x,y,z$), where  $\sigma_{x,y,z}$ are Pauli matrices and $|\psi_{\textbf{\textit{k}}} \rangle $ is wavefunction directly obtained from noncolinear DFT run. Spin textures for the four lowermost conduction bands around the K point are shown in Fig.~\ref{s1}. Spins align parallel or antiparallel to the $x$ direction and are independent of the \textbf{\textit{k}}, known as persistent spin texture (PST). These PSTs are protected by the mirror-symmetry ($\hat{M}_x$) and thus are not easily destroyed (see main text). These PSTs have the advantage of non-dephasing dissipationless spin transport and robustness towards spin-independent disorder~\cite{schliemann2017colloquium}. The degenerate bands CB and CB+1 have opposite spin configurations and segregate on different sublayers, as discussed in the main text. Similar observations are also seen for degenerate pair CB+2 and CB+3.
\newpage
\section{HZSS \MakeLowercase{for the valence bands around $\Gamma$ point}}
\begin{figure}[h]
	\includegraphics[width= 4 in]{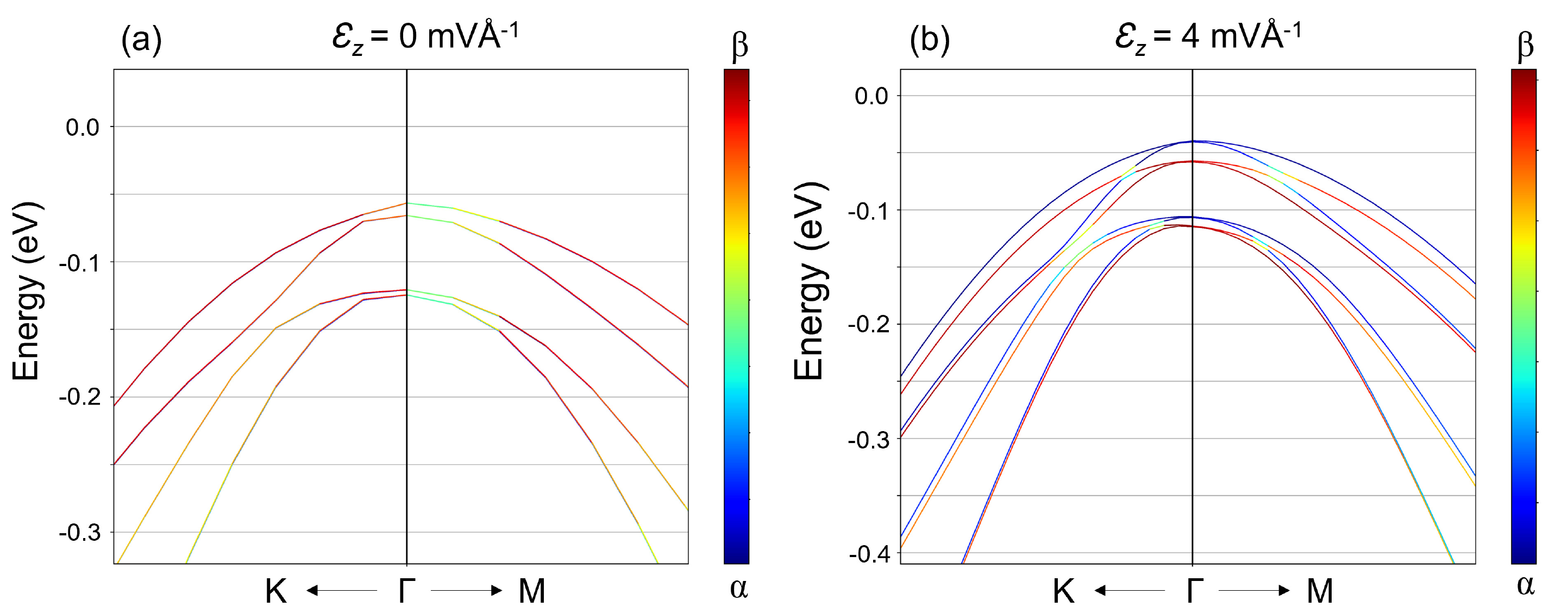}
	\caption{The layer-projected valence bands around the $\Gamma$ point in bilayer MnSe (a) without and (b) with an out-of-plane electric field ($\mathcal{E}_z$) of strength 4 mV\r{A}$^{-1}$. Here, $\alpha$ and $\beta$ represent the upper and lower layer, respectively. The layer-projected band structures are computed using the PyProcar~\cite{herath2020pyprocar}.}
	\label{s2}
\end{figure}

\begin{figure}[h]
	\includegraphics[width= 5 in]{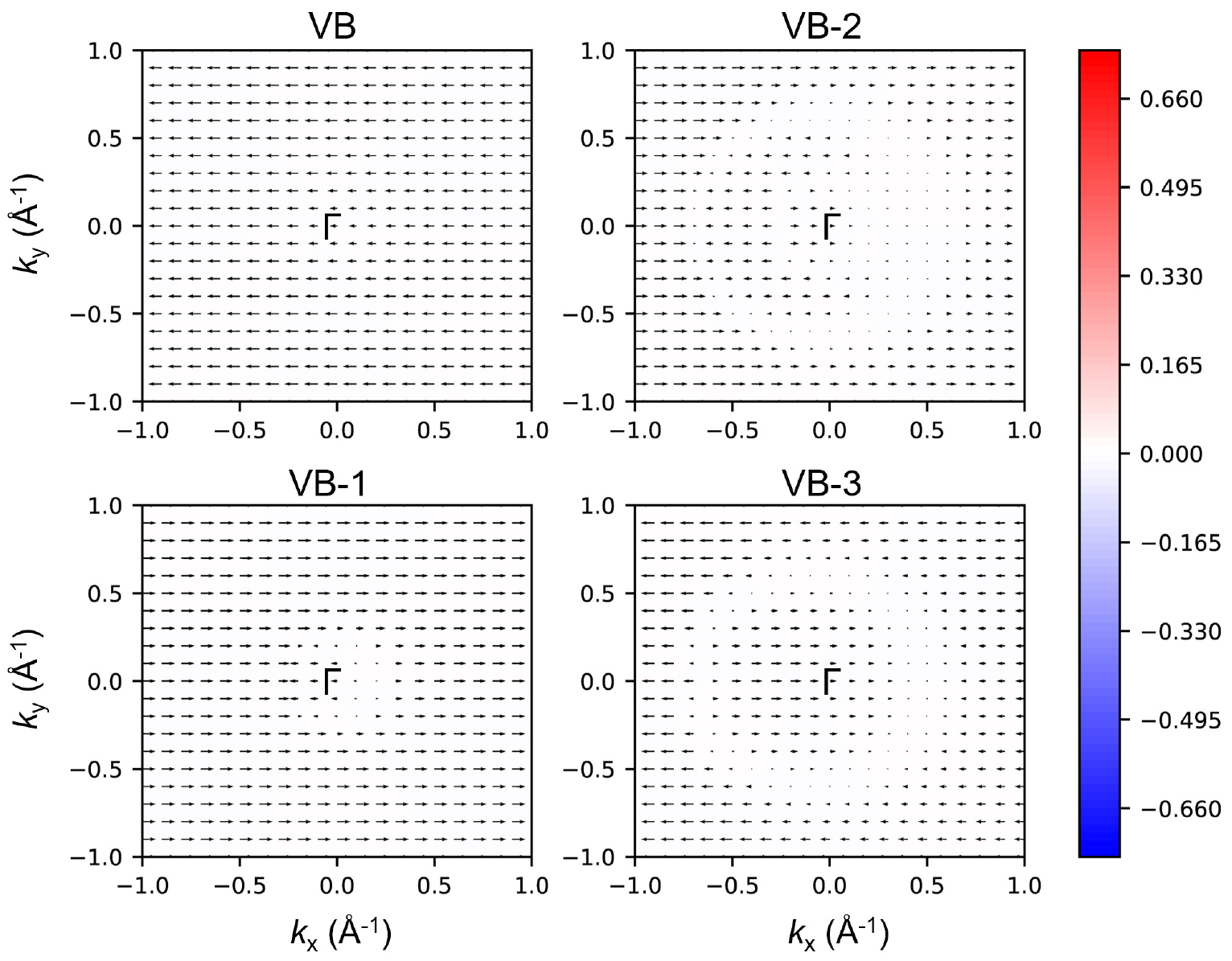}
	\caption{Spin textures of four highest valence bands VB, VB-1, VB-2, and VB-3 around the $\Gamma$ point in bilayer MnSe. The arrows and color represent the in-plane ($x,y$) and out-of-plane ($z$) components of spin textures, respectively.}
	\label{s3}
\end{figure}

As discussed in the main text, HZSS can also be observed for the VBs around the $\Gamma$ point in bilayer MnSe. We have plotted the layer-projected VBs in Fig.~\ref{s2}(a). Every band is doubly-degenerate due to $\mathcal{\hat{P}\hat{T}}$-symmetry. The degenerate pairs segregate on the different layers (not clearly visible in Fig.~\ref{s2}(a) since the red band overlaps the blue band). The degenerate bands have persistent spin textures with opposite orientations [Fig.~\ref{s3}]. The electric field along the $z$-direction ($\mathcal{E}_z$) breaks two-fold degeneracy by creating energy difference between $\alpha$ and $\beta$ bands. Figure~\ref{s2}(b) shows the band structure of top VBs around the $\Gamma$ point with $\mathcal{E}_z$ of strength 4 mV\r{A}$^{-1}$. The Zeeman splitting for the $\alpha$ bands is 72.1 meV, whereas $\beta$ bands are Zeeman split by 65.7 meV. Due to mirror-symmetry ($\hat{M}_x$), spins align along the $\pm$x direction [Fig.~\ref{s3}]. In a specific region, the orientations of the bands undergo a change, transitioning from the $+x$ direction to the $-x$ direction, or vice versa. These transitions occur at specific regions in $k$-space, where the energy levels of the bands become equal. Similar transitions in spin textures are also noticed in the previous works~\cite{zhao2020purely,bhattacharya2023spin}.

\section{E\MakeLowercase{lectric field reversal}}
\begin{figure}[h]
	\includegraphics[width= 5 in]{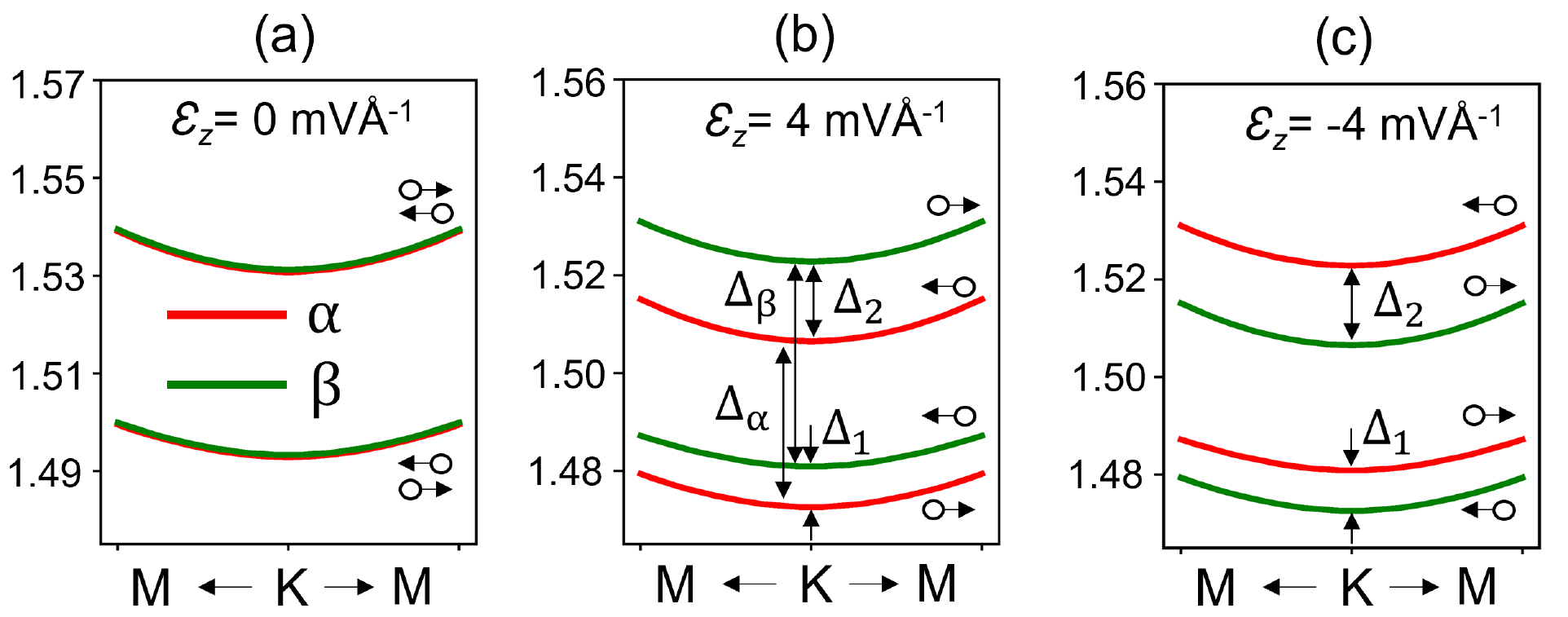}
	\caption{Layer-projected conduction bands around K point in bilayer MnSe (a) without and with $\mathcal{E}_z$ of strength 4 mV\AA$^{-1}$ along (b) $+z$ and (c) $-z$ direction, respetively. Here $\alpha$ and $\beta$ represent upper and lower layer, respectively.}
	\label{s4}
\end{figure}

In this section, we discuss the effect of electric field reversal on the band structure and spin textures of CBs around the K point in bilayer MnSe. We plot the layer-projected band structures structure without and with $\mathcal{E}_z$ of strength 4 mV{\AA}$^{-1}$ along the $+z$ and $-z$ direction [see Figs.~\ref{s4}(a),~\ref{s4}(b) and~\ref{s4}(c)]. As discussed in the main text, bands are twofold degenerate with a spin-sublayer locking effect. The $\mathcal{E}_z$ breaks the bands degeneracy by splitting the spin-up and spin-down bands. Reversing the direction of electric field from $\mathcal{E}_z$ to $-\mathcal{E}_z$ will switch the spin polarization [see Figs.~\ref{s4}(b) and~\ref{s4}(c)]. For instance, the lowest conduction band has spin orientation along the $+x$ and $-x$ directions with $\mathcal{E}_z$ along the $+z$ and $-z$ directions, respectively. In addition, the strength of splittings $|\Delta_1|$ and $|\Delta_2|$ remains the same in each case. However, $\Delta_i$ $(i=1,2)$ switch their sign under electric field reversal.

\section{E\MakeLowercase{xtension to Multilayer}}
\begin{figure}[htp]
	\includegraphics[width=5 in]{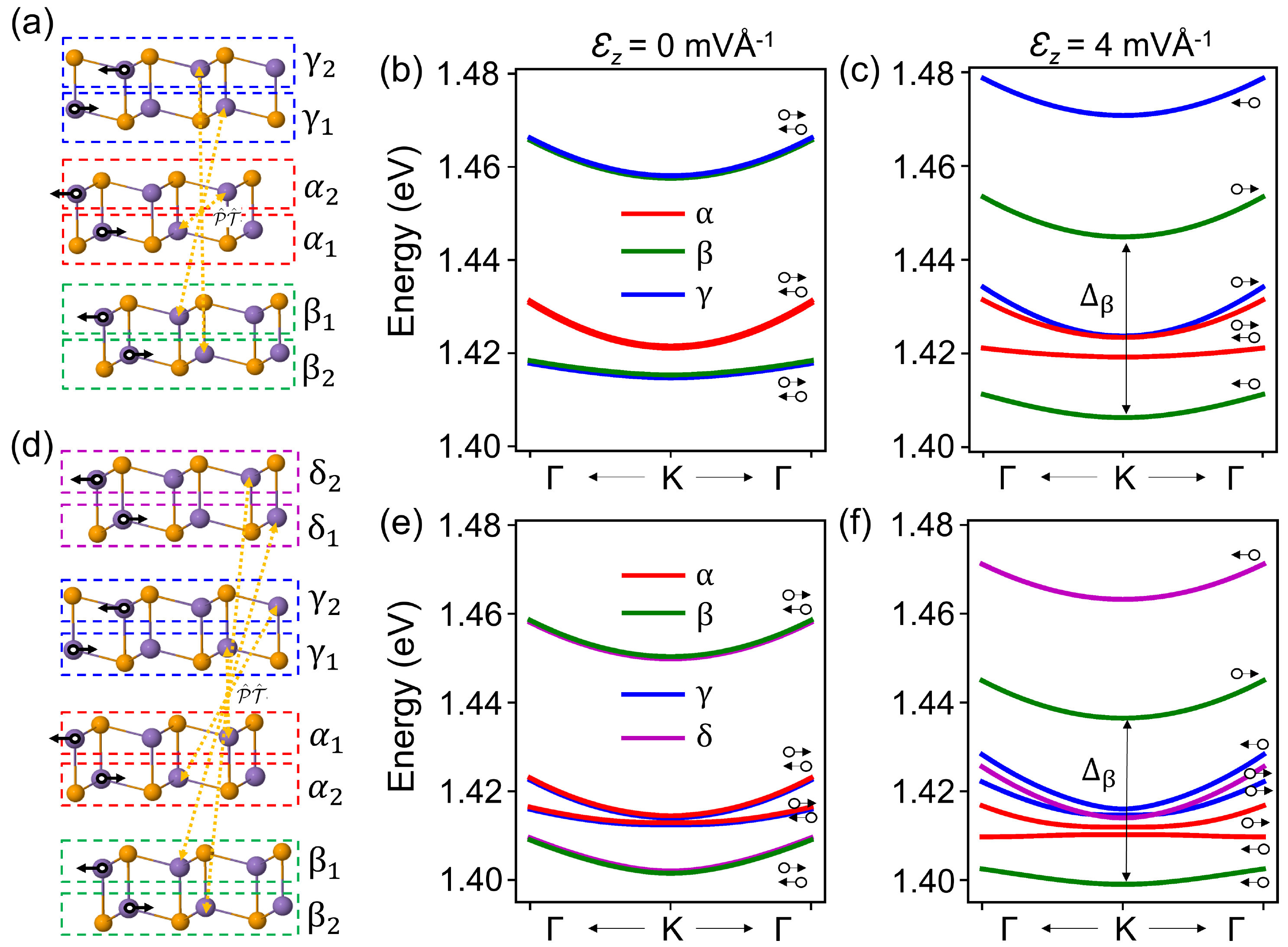}
	\caption{(a) The crystal structure of trilayer MnSe, in which every layer is divided into two sublayers. $\mathcal{\hat{P}\hat{T}}$ symmetry connects sublayers through orange dashed lines. The layer-projected conduction bands of trilayer MnSe (b) without and (c) with $\mathcal{E}_z$ of strength 4 mV\AA$^{-1}$. [(d)-(f)] are counterparts of [(a)-(c)] but for tetralayer MnSe. The most stable magnetic configurations for trilayer and tetralayer are $M$$\uparrow \downarrow \uparrow \downarrow \uparrow \downarrow$ and $M$$\uparrow \downarrow \uparrow \downarrow \uparrow \downarrow \uparrow \downarrow$, respectively.}
	\label{p5}
\end{figure}

In the realm of material synthesis, there has been a notable development in the creation of multilayer 2D materials, such as trilayers and tetralayers composed of graphene~\cite{lui2011observation, wirth2022experimental} and MoS$_2$~\cite{meng2022sliding,zhao2013interlayer}. The fundamental question of layer dependence in HZSS is still open and should now be attended to. Here, we consider the trilayer and tetralayer MnSe to check the layer dependence of results obtained for bilayer MnSe. Stacking orders are obtained similarly to bilayer MnSe, wherein the top layer is translated along the basal plane to achieve minimum energy stacking. The most stable stackings of trilayer and tetralayer are shown in Figs.~\ref{p5}(a) and~\ref{p5}(d), respectively. The MPG of trilayer and tetralayer MnSe is $2'/m$, and $\mathcal{\hat{P}\hat{T}}$ symmetry remains intact. Therefore, HZSS is also observed in trilayer and tetralayer MnSe, where every band is doubly degenerate, and degenerate pairs are localized on different sublayers with opposite spin magnetization [Figs.~\ref{p5}(b) and~\ref{p5}(e)]. To compare the strength of hidden splitting of trilayer and tetralayer with bilayer MnSe, we focus on the lowermost layer, i.e. $\beta$ layer in each case. $\Delta_{\beta}$ is observed to be 42.7 and 42.9 meV for trilayer and tetralayer MnSe, respectively, nearly similar to 39.5 meV for the case bilayer MnSe. The electric field along the $z$-direction splits the degenerate pairs, as shown in Figs.~\ref{p5}(c) and~\ref{p5}(f). With a tiny electric field of 4 mV\AA$^{-1}$, $\Delta_{\beta}$ changes to the 38.6 and 38.3 meV in the case of trilayer and tetralayer MnSe, respectively, both of which are comparable to those observed in the bilayer MnSe. Since MPG symmetry remains invariant with increasing layer number, similar results are also expected for the higher number of layers ($n>4$). Therefore, we infer that HZSS can be achieved in layered MnSe with arbitrary thickness. Thus the experimental realization of HZSS in layered MnSe is easily accessible. 

\section*{References}
\bibliography{ref}